\documentclass[twocolumn]{aastex631}
\usepackage{amsmath,bm,color}
\allowdisplaybreaks[2]
\begin{document}
\title{Convective Oscillations of Cosmic Rays}
\author{Yiran Zhang}
\affiliation{School of Physical Science and Technology, Southwest Jiaotong University, Chengdu 611756, China}
\author{Siming Liu}
\affiliation{School of Physical Science and Technology, Southwest Jiaotong University, Chengdu 611756, China}
\correspondingauthor{Yiran Zhang}
\email{zhangyr@swjtu.edu.cn}
\begin{abstract}
Turbulent motion of the background plasma can introduce fluctuations in the spatial distribution of cosmic rays (CRs). For an ensemble-averaged Gaussian CR distribution with a width $ \Lambda\left( t,p \right) $ at time $ t $ and momentum $ p $, it is found that $ P_\text{s}\left( t,k,p \right) $ and $ P_\text{d}\left( t,k,p \right) $, the fluctuation power spectra induced by incompressible and acoustic turbulent convection, respectively, have power-law asymptotics in wavenumber $ k $: $ P_\text{s}\left( t,k,p \right) \propto k^4 $ and $ P_\text{d}\left( t,k,p \right) \propto k^2 $ for $ k\ll 1/\Lambda\left( t,p \right) $; $ P_\text{s}\left( t,k,p \right) \propto k^{\gamma _\text{s}-4} $ for $ k\gg 1/L_\text{s} $, and $ P_\text{d}\left( t,k,p \right) \propto k^{\gamma _\text{d}-2} $ for $ k\gg 1/L_\text{d} $, where $ \gamma _\text{s} $ and $ \gamma _\text{d} $ are the spectral indices of the velocity power spectra of the incompressible and acoustic turbulence below their outer scales $ L_\text{s} $ and $ L_\text{d} $, respectively. These results have implications for probing interstellar turbulence via future high-resolution images of pulsar gamma-ray halos.
\end{abstract}
\section{Introduction}
Cosmic rays (CRs) are charged relativistic particles in cosmic space. It is generally considered that they are energized by extreme astrophysical sources, e.g., supernova remnants (SNRs) and pulsar wind nebulae \citep[PWNe;][]{2022RvMPP...6...19L}. Since their trajectories can be chaotized by interstellar magnetic field fluctuations, Galactic CRs may undergo diffusive propagation \citep{1966ApJ...146..480J}. Meanwhile, high-energy radiation can be generated during the interactions between the CRs and background fields, providing a window for detecting the CR distribution in deep space.

So far, the bulk energy spectra of the radiation from individual sources have been studied extensively \citep{2019ApJ...874...50Z,2024RAA....24g5016L}, while determining the spatial distribution of the emitting CRs remains a challenge. This difficulty partly arises from the unknown background fields; for example, synchrotron, bremsstrahlung, and hadronic radiation depend on the background magnetic field and gas, which can be highly nonuniform \citep{2008A&A...481..401A,2026A&A...708A.276C}. On the other hand, CR leptons can also emit gamma rays via inverse Compton (IC) scattering off background low-energy photons, whose spatial distribution around a single source should be nearly uniform \citep{2011ApJ...727...38S}. Thus, the near-source distribution of the IC radiation may have a similar spatial dependence to that of the leptons, rendering PWNe, which produce mainly electrons and positrons, the ideal tracers of the CR distribution morphology. However, existing gamma-ray images observed around individual PWNe or SNRs are still too blurred because of insufficient statistical significance \citep{2017Sci...358..911A}.

Nevertheless, recent observations of PWN gamma-ray halos do show hints of asymmetric morphologies \citep{2024ApJ...974..246A,2025SCPMA..6879504C,2025arXiv251006786C}. It can be expected that future precise measurements would reveal more irregular spatial structures of such halos. Some work has attempted to model the asymmetric halos with anisotropic diffusion of CRs \citep{2019PhRvL.123v1103L,2022ApJ...935...65Y,2026PhRvD.113b1303L,2026ApJ..1003..149W}. However, the diffusion equation has no irregular solution unless one inputs an irregular diffusion coefficient \citep{2026MNRAS.545f2108E}, which is beyond the classical theory, but may exist as a result of intermittent magnetic structures \citep{2023PhRvE.107f5206T,2023MNRAS.525.4985K,2024ApJ...975...65Z,2025PhRvL.135s1002B}.

In the classical theory of macroscopic transport, CRs can interact with large-scale turbulence via convection. It has been shown by early research that random compression is one of the mechanisms for stochastic acceleration \citep{1988SvAL...14..255P,1993PhyU...36.1020B,2010ApJ...713..475J}. Recent studies have also suggested that turbulent convection can produce small-scale anisotropies in the CR momentum distribution \citep{2022ApJ...938..106Z,2024ApJ...964L...1Z,2025ApJ...988...20Z}. In principle, such convection should also be able to generate random anisotropies in position space, serving as a possible origin of the spatial irregularities of the PWN halos.

In this paper, we investigate the convective fluctuations of the CR spatial distribution near a CR source. A seemingly related but distinct study was carried out by \cite{2023ApJ...951...38C}. The authors argued that the power spectrum of CR density fluctuations has a spectral index of $ -8/3 $ in the small-scale limit. However, they assumed that CRs constitute an isolated system, in conflict with the conventional picture. Another limitation of their model for our problem of interest lies in the exclusion of radiative energy losses. To the best of our knowledge, no other theoretical work has studied the CR spatial spectrum.

Regarding observations, the near-Earth CR power spectrum in the frequency domain has been measured by ground-based neutron monitors and muon detectors \citep{2016ScChE..59..547K,2019JGRA..124..804V}. However, such observations represent temporal responses of low-energy CRs to solar modulation, and cannot reveal the true spatial spectrum of interest for high-energy CRs whose Larmor radii are much larger than the heliospheric scale. Although one expects that the spatial spectrum of TeV CRs in deep space can be derived from gamma-ray images of PWN halos, there is currently a lack of such analysis. Our work will provide theoretical predictions for future observations.
\section{Convective Oscillations}
We follow the common view that CRs are scattered by magnetic field fluctuations in the background such that the CR system is not isolated. In general, the background field is also influenced by the CRs, particularly by low-energy CRs. Since we are interested in high-energy CRs whose energy density ($ \sim 0.01\ \text{eV}/\text{cm}^3 $ when integrated over $ cp>\text{TeV} $) is negligible compared with the background pressure ($ \sim\text{eV}/\text{cm}^3 $), a test particle theory can be developed for simplicity.

One of the central assumptions of diffusion is that particles forget their initial states after being scattered. This picture entails convection, the collective comovement of the particles with scattering centers. It is well known that nonuniform convection may lead to compression or expansion, with which the particles are adiabatically accelerated or decelerated, respectively. Meanwhile, the particles can also be decelerated because of radiation.

The above processes are encoded in the following convection--diffusion equation
\begin{align}
\frac{\partial f}{\partial t}&+\nabla \cdot \left( \boldsymbol{u}f-\mathbf{K}\cdot \nabla f \right) \notag\\
&-\frac{1}{p^2}\frac{\partial}{\partial p}\left\{ p^2\left[ \left( \frac{p}{3}\nabla \cdot \boldsymbol{u}+b \right) f+D\frac{\partial f}{\partial p} \right] \right\} =Q\label{PE}
\end{align}
\citep{1965P&SS...13....9P}, where $ t $ is the evolving time, $ f $ is the distribution function averaged over momentum directions, $ \boldsymbol{u} $ is the convection velocity, $ b $ is the loss rate of the momentum magnitude $ p $, $ \mathbf{K} $ is the spatial diffusion tensor due to wave--particle interactions \citep{2000PhRvL..85.4656C}, $ D $ is the momentum diffusion coefficient due to viscosity \citep{1988ApJ...331L..91E} and the wave--particle interactions, and $ Q $ is the particle injection rate per unit phase-space volume. For test particles, $ \boldsymbol{u} $, $ \mathbf{K} $, $ D $, $ b $, and $ Q $ are independent of $ f $.

Note that $ \mathbf{K} $ and $ D $ do not include the effect of so-called turbulent diffusion \citep{1988SvAL...14..255P,1993PhyU...36.1020B}, which can be derived with a further ensemble average of Equation \eqref{PE} over realizations of $ \boldsymbol{u} $. On the other hand, $ \mathbf{K} $, $ D $, and $ b $ will be assumed to be (temporally and spatially) uniform, consistent with the classical picture of CR propagation on PWN halo scales. In addition, following the conventional view that re-acceleration does not dominate the propagation process, $ D $ will be set to zero for further simplicity.

We consider that the overall distribution $ f_0 $ is controlled by the injection--diffusion--loss process without mean convection, and there is a fluctuating distribution $ f_1=f-f_0 $ due to a perturbation convection field $ \boldsymbol{u}_1=\boldsymbol{u} $. Then, Equation \eqref{PE} can be linearized as
\begin{align}
\frac{\partial f_0}{\partial t}-\mathbf{K}:\nabla \nabla f_0-\frac{1}{p^2}\frac{\partial}{\partial p}\left( p^2bf_0 \right) &=Q,\label{f0Eq}\\
\frac{\partial f_1}{\partial t}-\mathbf{K}:\nabla \nabla f_1-\frac{1}{p^2}\frac{\partial}{\partial p}\left( p^2bf_1 \right) &=\frac{p\nabla \cdot \boldsymbol{u}_1}{3}\frac{\partial f_0}{\partial p}\notag\\
&\quad -\boldsymbol{u}_1\cdot \nabla f_0,\label{f1Eq}
\end{align}
where the “:” symbol between two dyadics denotes the double-dot product.

We are interested in the near-source solution of Equation \eqref{f0Eq}, whose spatial dependence cannot be neglected. Hence, the solution of interest of Equation \eqref{f1Eq} is statistically inhomogeneous. Theoretically, it is possible to utilize the eikonal approximation \citep{1975ctf..book.....L} to define a locally quasi-homogeneous random field. However, gamma-ray observations of PWN halos may not have enough resolution to perform the eikonal analysis. A more feasible approach may be based on the global Fourier decomposition
\begin{align}
\left[ \begin{matrix}
f_0\left( t,\boldsymbol{r},p \right) \\
f_1\left( t,\boldsymbol{r},p \right) \\
\boldsymbol{u}_1\left( t,\boldsymbol{r} \right) 
\end{matrix} \right] =\int{\left[ \begin{matrix}
\tilde{f}_0\left( t,\boldsymbol{k},p \right) \\
\tilde{f}_1\left( t,\boldsymbol{k},p \right) \\
\tilde{\boldsymbol{u}}_1\left( t,\boldsymbol{k} \right) 
\end{matrix} \right] e^{i\boldsymbol{k}\cdot \boldsymbol{r}}\frac{d^3k}{\left( 2\pi \right) ^3}},\label{FE}
\end{align}
where the wave vector $ \boldsymbol{k} $ (together with $ \tilde{f}_0 $, $ \tilde{f}_1 $, and $ \tilde{\boldsymbol{u}}_1 $) is independent of the position vector $ \boldsymbol{r} $.

Substituting Equation \eqref{FE} into Equation \eqref{f1Eq}, it is not difficult to find that the nontrivial solution must match the condition $ \boldsymbol{k}=\boldsymbol{k}_0+\boldsymbol{k}_1 $ for $ \tilde{f}_0\left( t,\boldsymbol{k}_0,p \right) $, $ \tilde{f}_1\left( t,\boldsymbol{k},p \right) $, and $ \tilde{\boldsymbol{u}}_1\left( t,\boldsymbol{k}_1 \right) $. This ``three-wave coupling'' renders
\begin{align}
&\left[ \frac{\partial}{\partial t}-b\left( p \right) \frac{\partial}{\partial p}+\mathbf{K}\left( p \right) :\boldsymbol{kk} \right] \left[ p^2b\left( p \right) \tilde{f}_1\left( t,\boldsymbol{k},p \right) \right] \notag\\
&=-p^2b\left( p \right) g\left( t,\boldsymbol{k},p \right) ,\label{f1kEq}\\
&g\left( t,\boldsymbol{k},p \right) =i\int\frac{d^3k_1}{\left( 2\pi \right) ^3}\tilde{\boldsymbol{u}}_1\left( t,\boldsymbol{k}_1 \right) \notag\\
&\hspace{4em}\cdot \left[ \boldsymbol{k}\tilde{f}_0\left( t,\boldsymbol{k}-\boldsymbol{k}_1,p \right) -\boldsymbol{k}_1h\left( t,\boldsymbol{k}-\boldsymbol{k}_1,p \right) \right] ,\label{gk}\\
&h\left( t,\boldsymbol{k},p \right) =\tilde{f}_0\left( t,\boldsymbol{k},p \right) +\frac{1}{3}\frac{\partial \tilde{f}_0\left( t,\boldsymbol{k},p \right)}{\partial \ln p}.
\end{align}

It is because of the assumption $ D=0 $ that Equation \eqref{f1kEq} follows the Lagrangian derivative $ d/dt=\partial /\partial t-b\left( p\right) \partial /\partial p $ (with $ b\left( p\right) =-dp/dt $). Thus, the equation can readily be integrated over $ t_0 $ along the characteristic curve
\begin{align}
t-t_0=\tau \left( p,p_0 \right) =\int_p^{p_0}{\frac{dp'}{b\left( p' \right)}}\label{tau},
\end{align}
where $ \tau \left( p,p_0 \right) $ is the time for a particle losing its momentum from $ p_0 $ to $ p $ \citep{2010A&A...524A..51D}. Alternatively, the solution can also be expressed as an integral over $ p_0 $,
\begin{align}
p^2b\left( p\right) \tilde{f}_1\left( t,\boldsymbol{k},p \right) &=\int_{p_{\text{m}}\left( t,p\right)}^pg\left[ t-\tau \left( p,p_0 \right) ,\boldsymbol{k},p_0 \right] \notag\\*
&\quad\times e^{\boldsymbol{k}\boldsymbol{k}:\int_{p_0}^p{\frac{\boldsymbol{K}\left( p_1 \right)}{b\left( p_1 \right)}dp_1}}p_{0}^{2}dp_0,\label{f1k}
\end{align}
where $ p_{\text{m}}\left( t,p\right) $ is determined by $ \tau\left( p,p_\text{m} \right) =t $, and we have taken the initial condition $ \tilde{f}_1\left( 0,\boldsymbol{k},p \right) =0 $. Physically, we need to impose $ p_{\text{m}}\left( t,p\right) \rightarrow\infty $ for $ t>t_\text{c}\left( p \right) $, where $ t_\text{c}=\tau\left( p,\infty \right) $.

The statistics of the random field $ f_1 $ are encoded in the correlation function $ \left< \tilde{f}_1\left( t,\boldsymbol{k},p \right) \tilde{f}_1^*\left( t,\boldsymbol{k}',p \right) \right> $, where the angle brackets represent the ensemble average over realizations of convection. For simplicity, we consider a homogeneous isotropic configuration for the statistics of the convection field \citep{1988SvAL...14..255P},
\begin{align}
&\left< \tilde{\boldsymbol{u}}_1\left( t,\boldsymbol{k} \right) \tilde{\boldsymbol{u}}_1^*\left( t',\boldsymbol{k}' \right) \right> =\left( 2\pi \right) ^6\delta ^3\left( \boldsymbol{k}-\boldsymbol{k}' \right) \notag\\
&\hspace{-1em}\times \left[ \frac{W_{\text{s}}\left( t-t',k \right)}{2}\left( \mathbf{I}-\frac{\boldsymbol{k}\boldsymbol{k}}{k^2} \right) +W_{\text{d}}\left( t-t',k \right) \frac{\boldsymbol{k}\boldsymbol{k}}{k^2} \right] ,\label{u1ku1k'}
\end{align}
which may coincide with the correlation function of the background flow field in the test particle regime. Here, $ W_\text{s} $ and $ W_\text{d} $ are the solenoidal (incompressible) and dilatational (acoustic) components of the Fourier-space power density of the turbulent velocity field, respectively, $ \delta ^3\left( \boldsymbol{k}-\boldsymbol{k}' \right) $ is the 3D Dirac delta function, and $ \mathbf{I} $ is the identity tensor. For convenience, we will sometimes use the subscript $ j $ to denote the s or d component.

From Equations \eqref{gk}, \eqref{f1k}, and \eqref{u1ku1k'}, we have
\begin{widetext}
\begin{align}
\left< \tilde{f}_1\left( t,\boldsymbol{k},p \right) \tilde{f}_{1}^{*}\left( t' ,\boldsymbol{k}' ,p' \right) \right> &=\frac{1}{\left( pp' \right) ^2b\left( p \right) b\left( p' \right)}\int_{p_{\text{m}}\left( t,p \right)}^pdp_0\int_{p_{\text{m}}\left( t' ,p' \right)}^{p'}\left< g\left[ t-\tau \left( p,p_0 \right) ,\boldsymbol{k},p_0 \right] g^*\left[ t' -\tau \left( p' ,p_0' \right) ,\boldsymbol{k}' ,p_0' \right] \right> \notag\label{f1kf1k'}\\
&\quad\times e^{\boldsymbol{kk}:\int_{p_0}^p{\frac{\boldsymbol{K}\left( p_1 \right)}{b\left( p_1 \right)}dp_1}+\boldsymbol{k}' \boldsymbol{k}' :\int_{p_0'}^{p'}{\frac{\boldsymbol{K}\left( p_1 \right)}{b\left( p_1 \right)}dp_1}}\left( p_0p_0' \right) ^2dp_0' ,\\
\left< g\left( t,\boldsymbol{k},p \right) g^*\left( t',\boldsymbol{k}',p' \right) \right> &=G_{\text{s}}\left( t,t';\boldsymbol{k},\boldsymbol{k}';p,p' \right) +G_{\text{d}}\left( t,t';\boldsymbol{k},\boldsymbol{k}';p,p' \right) ,\\
G_{\text{s}}\left( t,t';\boldsymbol{k},\boldsymbol{k}';p,p' \right) &=\frac{1}{2}\int{W_{\text{s}}\left( t-t',k_1 \right) \left( \mathbf{I}-\frac{\boldsymbol{k}_1\boldsymbol{k}_1}{k_1^2} \right) :\boldsymbol{k}\boldsymbol{k}'\tilde{f}_0\left( t,\boldsymbol{k}-\boldsymbol{k}_1,p \right) \tilde{f}_0^*\left( t',\boldsymbol{k}' -\boldsymbol{k}_1,p' \right) d^3k_1},\label{Ctkk'}\\
G_{\text{d}}\left( t,t';\boldsymbol{k},\boldsymbol{k}';p,p' \right) &=\int W_{\text{d}}\left( t-t',k_1 \right) \frac{\boldsymbol{k}_1\boldsymbol{k}_1}{k_1^2}\notag\\
&\hspace{-2em}:\left[ \boldsymbol{k}\tilde{f}_0\left( t,\boldsymbol{k}-\boldsymbol{k}_1,p \right) -\boldsymbol{k}_1h\left( t,\boldsymbol{k}-\boldsymbol{k}_1,p \right) \right] \left[ \boldsymbol{k}' \tilde{f}_0\left( t',\boldsymbol{k}' -\boldsymbol{k}_1,p' \right) -\boldsymbol{k}_1h\left( t',\boldsymbol{k}' -\boldsymbol{k}_1,p' \right) \right] ^*d^3k_1.\label{Clkk'}
\end{align}
\end{widetext}

In general, the two-point correlation determined by Equation \eqref{f1kf1k'} is non-diagonal and depends on the directions of $ \boldsymbol{k} $ and $ \boldsymbol{k}' $, corresponding to an inhomogeneous anisotropic random field. These asymmetries arise from the nonuniformity of $ f_0 $. If $ f_0 $ is spherically symmetric, the system has rotational symmetry such that $ f_1 $ is statistically spatially isotropic. For a uniform system $ \tilde{f}_0\left( t,\boldsymbol{k},p \right) =\left( 2\pi \right) ^3\delta ^3\left( \boldsymbol{k} \right) f_0\left( t,\boldsymbol{r},p \right) $, one has
\begin{align}
\left< g\left( t,\boldsymbol{k},p \right) g^*\left( t',\boldsymbol{k}',p' \right) \right> &=\left( 2\pi \right) ^6\delta ^3\left( \boldsymbol{k}-\boldsymbol{k}' \right) \notag\\
&\hspace{-10em}\times W_{\text{d}}\left( t-t',k \right) \left( \frac{k}{3} \right) ^2\frac{\partial f_0\left( t,\boldsymbol{r},p \right)}{\partial \ln p}\frac{\partial f_0\left( t',\boldsymbol{r},p' \right)}{\partial \ln p'},\label{gkgk'HI}
\end{align}
with which $ f_1 $ reduces to a homogeneous isotropic random field. This result also implies that small-scale fluctuations of a smooth nonuniform CR distribution are statistically quasi-homogeneous and likely to be of acoustic origin.
\section{CR Power Spectrum}
For simplicity, we assume the bulk distribution $ f_0 $ to be spherically symmetric. According to Equation \eqref{f0Eq}, this assumption is valid for spherically symmetric injection $ Q\left( t,\boldsymbol{r},p \right) =Q\left( t,r,p \right) $ and isotropic diffusion $ \mathbf{K}=\kappa\mathbf{I} $. Furthermore, we consider the size of the injection source to be small, consistent with the general understanding of the halo problem for aged SNRs and PWNe. The corresponding spatial morphology of $ f_0 $ should be Gaussian-like. Thus, we take
\begin{align}
f_0\left( t,\boldsymbol{r},p \right) &=\frac{a\left( t,p \right)}{\left[ \sqrt{2\pi}\Lambda \left( t,p \right) \right] ^3}e^{-\frac{1}{2}\left[ \frac{r}{\Lambda \left( t,p \right)} \right] ^2},\label{f0}\\
\tilde{f}_0\left( t,\boldsymbol{k},p \right) &=a\left( t,p \right) e^{-\frac{\left[ \Lambda \left( t,p \right) k \right] ^2}{2}}.\label{f0k}
\end{align}

Equations \eqref{f0} and \eqref{f0k} are exact in the case of burst point injection $ Q\left( t,\boldsymbol{r},p \right) =N\left( p \right) \delta \left( t \right) \delta ^3\left( \boldsymbol{r} \right) $, which yields
\begin{align}
a\left( t,p \right) &=\frac{b\left[ p_{\text{m}}\left( t,p \right) \right]}{b\left( p \right)}N\left[ p_{\text{m}}\left( t,p \right) \right] \Theta\left[ t_\text{c}\left( p\right) -t \right] ,\label{aSNR}\\
\Lambda ^2\left( t,p \right) &=\lambda ^2\left[ p,p_{\text{m}}\left( t,p \right) \right] =2\int_p^{p_{\text{m}}\left( t,p \right)}{\frac{\kappa \left( p' \right)}{b\left( p' \right)}dp'},\label{LambdaSNR}
\end{align}
with $ \Theta $ denoting the unit step function, $ N $ the injected number of particles per unit momentum-space volume, and $ \lambda\left( p,p_0 \right) $ the diffusion length scale of a particle propagating with energy losses for a time $ \tau\left( p,p_0 \right) $ \citep{2010A&A...524A..51D}. For other scenarios of injection, the Gaussian functions may be treated as approximations preserving the finiteness and localization properties of physical solutions: for example, the typical unphysical singularity $ f_0\left( t,\boldsymbol{r},p \right) \propto 1/r $ as $ r\rightarrow 0 $ in the continuous point injection model $ Q\left( t,\boldsymbol{r},p \right) =\dot{N}\left( t,p \right) \delta ^3\left( \boldsymbol{r} \right) $ can be regularized by the finite-size effect of the injection source such that $ \tilde{f}_0\left( t,\boldsymbol{k},p \right)  $ has a cutoff toward $ k\rightarrow\infty $; in addition, because of the energy losses and/or finite-time evolution, $ f_0\left( t,\boldsymbol{r},p \right) $ decays much faster than $ 1/r $ as $ r\rightarrow\infty $, leading to a finite $ \tilde{f}_0\left( t,\boldsymbol{0},p \right) $.

To calculate $ G_j\left( t,t';\boldsymbol{k},\boldsymbol{k}';p,p' \right) $, we introduce a coordinate system whose polar axis is aligned with $ \boldsymbol{k} $, and denote the (polar, azimuthal) angular coordinates of $ \boldsymbol{k}' $ and $ \boldsymbol{k}_1 $ in such a system as $ \left( \theta ',\varphi ' \right) $ and $ \left( \theta _1,\varphi _1 \right) $, respectively. Note that the angle $ \theta _1' $ between $ \boldsymbol{k}' $ and $ \boldsymbol{k}_1 $ is determined by $ \cos \theta _1'=\cos \theta ' \cos \theta _1+\sin \theta '\sin \theta _1\cos \left( \varphi '-\varphi _1 \right) $. Substituting Equation \eqref{f0k} into Equations \eqref{Ctkk'} and \eqref{Clkk'}, it is not difficult to find that the integrals over $ \varphi _1 $ can be expressed in terms of Bessel functions, and the results are independent of $ \varphi ' $. The fact that $ \theta ' $ is the only angular parameter allowed in $ \left< \tilde{f}_1\left( t,\boldsymbol{k},p \right) \tilde{f}_1^*\left( t,\boldsymbol{k}',p \right) \right> $ implies that $ f_1 $ is an inhomogeneous isotropic random field.

We are interested in the position-independent power spectrum $ P=P_\text{s}+P_\text{d} $ of this random field, where $ P_\text{s} $ and $ P_\text{d} $ are the component spectra associated with incompressible and acoustic turbulent convection, respectively. We define
\begin{align}
P\left( t,k,p \right) &=\mathcal{P}\left( t;k;p,p \right) \notag\\
&=4\pi k^2\left< \tilde{f}_1\left( t,\boldsymbol{k},p \right) \tilde{f}_{1}^{*}\left( t,\boldsymbol{k} ,p \right) \right>,
\end{align}
which gives the spatially integrated power per unit wavenumber. According to Equations \eqref{f1kf1k'}--\eqref{Clkk'}, one has
\begin{widetext}
\begin{align}
\mathcal{P} _j\left( t;k;p,p' \right) &=\frac{4\pi k^2}{\left( pp' \right) ^2b\left( p \right) b\left( p' \right)}\int_{p_{\text{m}}\left( t,p \right)}^pdp_0\int_{p_{\text{m}}\left( t,p' \right)}^{p'}G_j\left[ t-\tau \left( p,p_0 \right) ,t-\tau \left( p' ,p_0' \right) ;\boldsymbol{k},\boldsymbol{k};p_0,p_0' \right] \notag\\
&\quad\times e^{-\frac{k^2}{2}\left[ \lambda ^2\left( p,p_0 \right) +\lambda ^2\left( p' ,p_0' \right) \right]}\left( p_0p_0' \right) ^2dp_0' ,\label{Ptlk}\\
G_{\text{s}}\left( t,t' ;\boldsymbol{k},\boldsymbol{k};p,p' \right) &=\frac{2\pi a\left( t,p \right) a\left( t' ,p' \right)}{C^2\left( t,t' ;p,p' \right)}\int_0^{\infty}W_{\text{s}}\left( t-t' ,k_1 \right) e^{-\frac{C\left( t,t' ;p,p' \right)}{2}\left( k-k_1 \right) ^2}\notag\\
&\quad\times \left[ 1+e^{-2kk_1C\left( t,t' ;p,p' \right)}-\frac{1-e^{-2kk_1C\left( t,t' ;p,p' \right)}}{kk_1C\left( t,t' ;p,p' \right)} \right] dk_1,\label{Ctkk}\\
G_{\text{d}}\left( t,t' ;\boldsymbol{k},\boldsymbol{k};p,p' \right) &=\frac{2\pi a\left( t,p \right) a\left( t' ,p' \right)}{C^2\left( t,t' ;p,p' \right)}\int_0^{\infty}W_{\text{d}}\left( t-t' ,k_1 \right) e^{-\frac{C\left( t,t' ;p,p' \right)}{2}\left( k-k_1 \right) ^2}\notag\\
&\hspace{-8em}\times \left\{ \Phi \left( t,t' ;k,k_1;p,p' \right) \left[ 1+e^{-2kk_1C\left( t,t' ;p,p' \right)}-\frac{1-e^{-2kk_1C\left( t,t' ;p,p' \right)}}{kk_1C\left( t,t' ;p,p' \right)} \right] +\Psi \left( t,t' ;k,k_1;p,p' \right) \left[ 1-e^{-2kk_1C\left( t,t' ;p,p' \right)} \right] \right\} dk_1,\label{Clkk}\\
\Phi\left( t,t' ;k,k_1 ;p,p' \right) &=kk_1 C\left( t,t' ;p,p' \right) \left[ A\left( t;k,k_1 ;p \right) B\left( t' ,k_1 ,p' \right) +A\left( t' ;k,k_1 ;p' \right) B\left( t,k_1 ,p \right) \right] -2B\left( t,k_1 ,p \right) B\left( t' ,k_1 ,p' \right) ,\\
\Psi\left( t,t' ;k,k_1 ;p,p' \right) &=kk_1 C\left( t,t' ;p,p' \right) \left[ A\left( t;k,k_1 ;p \right) A\left( t' ;k,k_1 ;p' \right) +B\left( t,k_1 ,p \right) B\left( t' ,k_1 ,p' \right) \right] ,\\
A\left( t;k,k_1 ;p \right) &=\frac{k_1}{k}\left[ 1+\frac{\alpha \left( t,p \right) -\beta \left( t,p \right) \Lambda ^2\left( t,p \right) \left( k^2+k_1^2 \right)}{3} \right] ,\quad B\left( t,k_1,p \right) =\frac{2}{3}\beta \left( t,p \right) \left[ \Lambda \left( t,p \right) k_1 \right] ^2-1,\\
C\left( t,t' ;p,p' \right) &=\Lambda ^2\left( t,p \right) +\Lambda ^2\left( t' ,p' \right) ,\quad \alpha \left( t,p \right) =\frac{\partial \ln a\left( t,p \right)}{\partial \ln p},\quad \beta \left( t,p \right) =\frac{\partial \ln \Lambda \left( t,p \right)}{\partial \ln p}.
\end{align}
\end{widetext}

It has been suggested by early research that Eulerian time decorrelation of turbulence in the global rest frame (where $ \left< \boldsymbol{u} \right> =\boldsymbol{0} $) arises mainly from random sweeping by energy-containing eddies \citep{2017AnRFM..49...51H,2020PhRvR...2b3357P}. For isotropic turbulence (Equation \ref{u1ku1k'}), one has
\begin{align}
W_{\text{s}}\left( t,k \right) &=e^{-\frac{\left( ku_\text{r}t \right) ^2}{2}}\frac{w_{\text{s}}\left( k \right)}{4\pi k^2},\label{Wt}\\
W_{\text{d}}\left( t,k \right) &=e^{-ikVt-\frac{\left( ku_\text{r}t \right) ^2}{2}}\frac{w_{\text{d}}\left( k \right)}{4\pi k^2},\label{Wl}
\end{align}
where $ w_\text{s} $ and $ w_\text{d} $ are the incompressible and acoustic turbulence spectra, respectively, $ u_\text{r} $ is the root-mean-square velocity of the energy-containing eddies, and $ V $ is the speed of sound. For simplicity, we consider power-law spectra in the inertial range, with $ w_j\left( k\right) \propto k^{\gamma _j}\Theta \left( L_jk-1 \right) $, where $ \gamma _j $ is the cascade spectral index, and $ L_j $ is the generation scale of turbulence. For $ \gamma_{\text{s}}<-1 $, the sweeping speed can be evaluated as $ u_\text{r}=\sqrt{w_{\text{s}}\left( 1/L_{\text{s}} \right) /L_{\text{s}}} $. A phenomenological setting is $ \gamma _j=-5/3 $, as indicated by the Big Power Law observed in the background fluid \citep{1995ApJ...443..209A}.

Note that the particles under consideration are interstellar ultra-relativistic electrons and positrons producing synchrotron and IC radiation. In the low-energy limit, the two radiation processes contribute comparable energy loss rates. As the energy increases, the growth of the IC loss rate becomes suppressed by the Klein--Nishina (KN) effect, while the energy dependence of the synchrotron loss rate remains unchanged \citep{2011ApJ...727...38S}. Thus, over a wide energy range, the total loss rate can be approximated via
\begin{align}
b\left( p\right) =\frac{4}{3}\sigma _{\text{T}}U\left( \frac{p}{mc}\right) ^2,\label{b}
\end{align}
where $ U $ is the energy density of the effective background photon field, $ \sigma _{\text{T}} $ is the Thomson cross section, $ m $ is the rest mass of an electron, and $ c $ is the speed of light. It follows from Equations \eqref{tau} and \eqref{LambdaSNR} that
\begin{align}
\tau \left( p,p_0 \right) &=t_{\text{c}}\left( p \right) -t_{\text{c}}\left( p_0 \right) ,\\*
\lambda ^2\left( p,p_0 \right) &=2\frac{\kappa \left( p \right) t_{\text{c}}\left( p \right) -\kappa \left( p_0 \right) t_{\text{c}}\left( p_0 \right)}{1-\Delta},\\*
p_{\text{m}}\left( t,p \right) &=\frac{pt_{\text{c}}\left( p \right)}{t_{\text{c}}\left( p \right) -t},\quad t_{\text{c}}\left( p \right) =\frac{p}{b\left( p \right)},\label{pm}
\end{align}
where we have assumed power-law energy scaling $ \kappa\left( p\right) \propto p^\Delta $ for diffusion \citep{2026SciA...12v8173T}.
\begin{figure*}
\includegraphics[width=\textwidth]{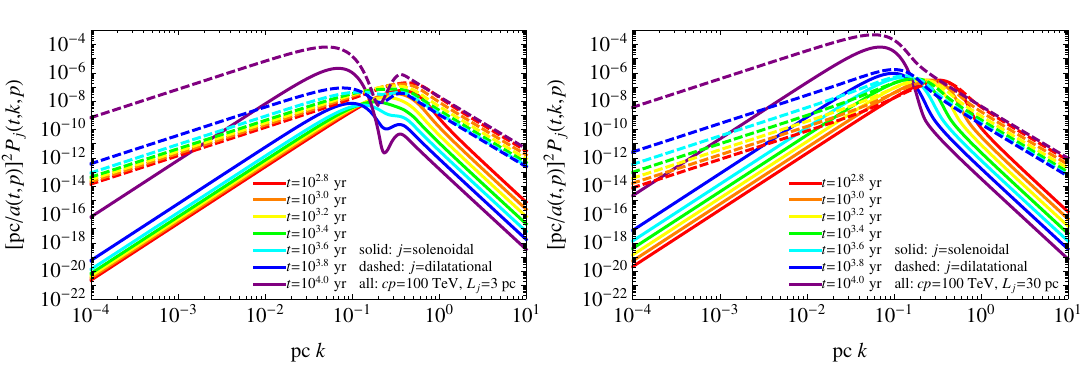}
\caption{Power spectra of convective oscillations for electrons and positrons around an SNR-like source.}\label{modSNR}
\end{figure*}
\begin{table*}
\centering
\caption{Fixed Parameters for Figures \ref{modSNR}--\ref{rad}.}\label{para}
\vspace{-\baselineskip}
\begin{tabular}{ccccccccc}
\hline
\hline
$ \partial\ln N/\partial\ln p $ & $ \kappa\left( \text{TeV}/c \right) $ & $ \Delta $ & $ V $ & $ \sqrt{w_{\text{s}}\left( 1/L_{\text{s}} \right) /L_{\text{s}}} $ & $ \sqrt{w_{\text{d}}\left( 1/L_{\text{d}} \right) /L_{\text{d}}} $ & $ \gamma _j $ & $ U $ & $ t_\text{b} $\\
\hline
$ -4 $ & $ 10^{27}\text{ cm}^2/\text{s} $ & 1/3 & 100 km/s & 10 km/s & $ 10/\sqrt{2} $ km/s & $ -5/3 $ & 0.3 $ \text{eV}/\text{cm}^3 $ & 10 kyr\\
\hline
\end{tabular}
\tablecomments{Equipartition of turbulence energy among spatial directions ($ w_\text{s}=2w_\text{d} $) has been assumed if $ L_{\text{s}}=L_{\text{d}} $.\\}
\end{table*}

The desired spectra of CR convective oscillations around a Gaussian distribution are given by Equations \eqref{Ptlk}--\eqref{pm}. In Figure \ref{modSNR}, we plot the spectra with Equations \eqref{aSNR} and \eqref{LambdaSNR} for 100 TeV electrons and positrons, whose interactions with the 2.7 K cosmic microwave background (CMB) produce mainly approximately 30 TeV photons. The fixed parameters can be found in Table \ref{para}. Since Figure \ref{modSNR} utilizes the burst injection model, the results are more likely to be related to an SNR-like source; efficient CR acceleration in the free expansion phase may be characterized as burst injection.

\begin{figure*}
\includegraphics[width=\textwidth]{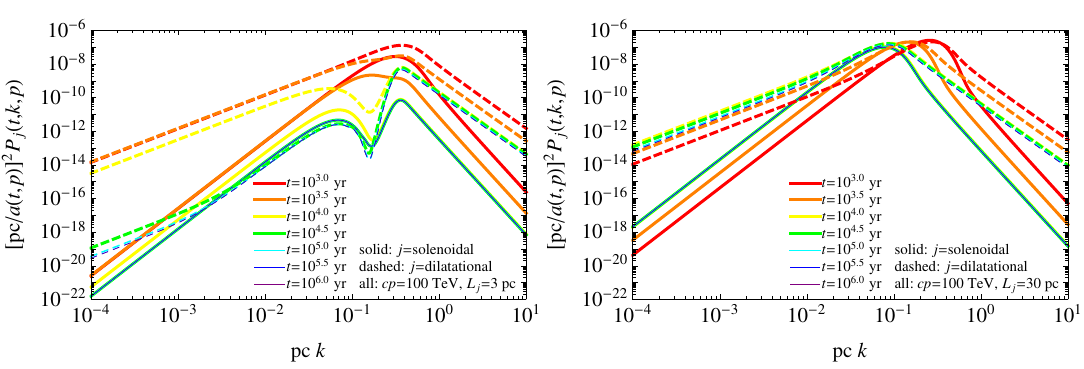}
\caption{Power spectra of convective oscillations for electrons and positrons around a PWN-like source. Note that the green, cyan, blue, and purple solid lines mostly overlap each other.}\label{modPWN}
\end{figure*}

\begin{figure}
\includegraphics[width=\columnwidth]{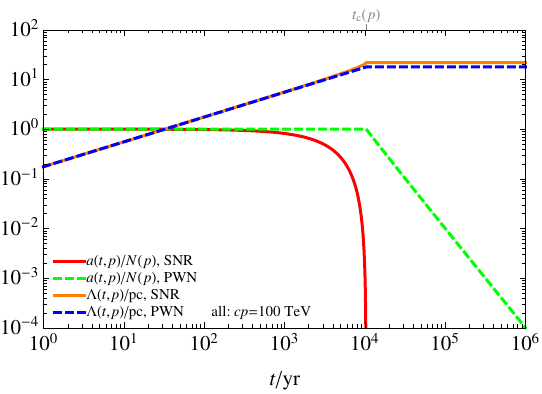}
\caption{Evolution of the amplitudes and widths of the modeled electron and positron halos for SNR- (Equations \ref{aSNR} and \ref{LambdaSNR}) and PWN-like (Equations \ref{aPWN} and \ref{LambdaPWN}) sources.}\label{a&Lambda}
\end{figure}

On the other hand, it is usually considered that PWN halos arise from continuous CR injection \citep{2010A&A...524A..51D}, which enables nontrivial solutions ($ f\ne 0 $) for $ t\geqslant t_\text{c} $. Although, as mentioned, the Gaussian distribution cannot be exactly achieved in the continuous injection model, it may encode the true limiting behaviors determining the large- and small-scale asymptotics of a physical power spectrum. In this sense, Equations \eqref{Ptlk}--\eqref{pm} provide a global approximation for the fluctuation spectra of the PWN CR halo. A phenomenological construction of the Gaussian halo can be given by
\begin{align}
a\left( t,p \right) &=N\left( p \right) \left\{ 1+\frac{t-\min \left[ t,t_{\text{c}}\left( p \right) \right]}{t_{\text{b}}} \right\} ^{-2},\label{aPWN}\\
\Lambda ^2\left( t,p \right) &=2\kappa \left( p \right) \min \left[ t,t_{\text{c}}\left( p \right) \right] ,\label{LambdaPWN}
\end{align}
where we have taken the following effects into account: (i) the pulsar spin-down luminosity is proportional to $ \left( 1+t/t_{\text{b}} \right) ^{-2} $ due to magnetic dipole radiation; (ii) CRs are accelerated at a constant efficiency such that their injection is proportional to the spin-down luminosity \citep{1995A&A...294L..41A}; (iii) the CRs undergo retarded propagation for a mean time $ \min \left( t,t_{\text{c}} \right) $. The fluctuation spectra of this halo are shown in Figure \ref{modPWN}. We also notice from Figure \ref{a&Lambda} that Equation \eqref{LambdaPWN} is just an analogy of Equation \eqref{LambdaSNR}.

As seen, Figures \ref{modSNR} and \ref{modPWN} imply that small-scale fluctuations of a physical CR halo tend to be dominated by acoustic oscillations (see also Equation \ref{gkgk'HI}). For $ k\rightarrow\infty $, the integrand in Equation \eqref{Ptlk} peaks around $ \left( p_0,p_0' \right) =\left( p,p' \right) $, and thus $ P_j\left( t,k,p \right) \propto k^{-2}G_j\left( t,t;\boldsymbol{k},\boldsymbol{k};p,p \right) $; meanwhile, the integrands in Equations \eqref{Ctkk} and \eqref{Clkk} peak around $ k_1=k $, for which $ G_{\text{s}}\left( t,t;\boldsymbol{k},\boldsymbol{k};p,p \right) \propto W_{\text{s}}\left( 0,k \right) $ and $ G_{\text{d}}\left( t,t;\boldsymbol{k},\boldsymbol{k};p,p \right) \propto k^2W_{\text{d}}\left( 0,k \right) $. Consequently, the small-scale asymptotic spectra arise from the coupling of $ \tilde{f}_0\left( t,\boldsymbol{0},p \right) $ and $ \tilde{\boldsymbol{u}}_1\left( t,\boldsymbol{k}_1 \right) $; with Equations \eqref{Wl} and \eqref{Wt}, they follow
\begin{align}
P_{\text{s}}\left( t,k,p \right) \propto k^{\gamma _{\text{s}}-4},\quad P_{\text{d}}\left( t,k,p \right) \propto k^{\gamma _{\text{d}}-2}.\label{SCL}
\end{align}
It should be emphasized that the only nonzero solution allowed by a uniform source ($ \nabla f_0=\boldsymbol{0} $) is $ P_\text{d}\left( t,k,p\right) $ for $ k\geqslant 1/L_\text{d} $.

Figures \ref{modSNR} and \ref{modPWN} also show that the halo can have significant fluctuations on scales larger than the cutoff scale of turbulence. For $ k\rightarrow 0 $, the exponential kernel in the integrand in Equation \eqref{Ptlk} no longer contributes a sharp peak in the $ p_0 $--$ p_0' $ plane; meanwhile, the integrands in Equations \eqref{Ctkk} and \eqref{Clkk} peak around $ k_1=1/L_\text{s} $ and $ k_1=1/L_\text{d} $, respectively, leading to $ G_{\text{s}}\left( t,t' ;\boldsymbol{k},\boldsymbol{k};p,p' \right) \propto k^2 $ and $ G_{\text{d}}\left( t,t' ;\boldsymbol{k},\boldsymbol{k};p,p' \right) \propto k^0 $. Thus, the large-scale asymptotic spectra follow
\begin{align}
P_{\text{s}}\left( t,k,p \right) \propto k^4,\quad P_{\text{d}}\left( t,k,p \right) \propto k^2,\label{LSL}
\end{align}
corresponding to the coupling of $ \tilde{f}_0\left[ t,-\boldsymbol{k}_1/\left( L_jk_1 \right) ,p \right] $ and $ \tilde{\boldsymbol{u}}_1\left[ t,\boldsymbol{k}_1/\left( L_jk_1 \right) \right] $.
\begin{figure*}
\includegraphics[width=\textwidth]{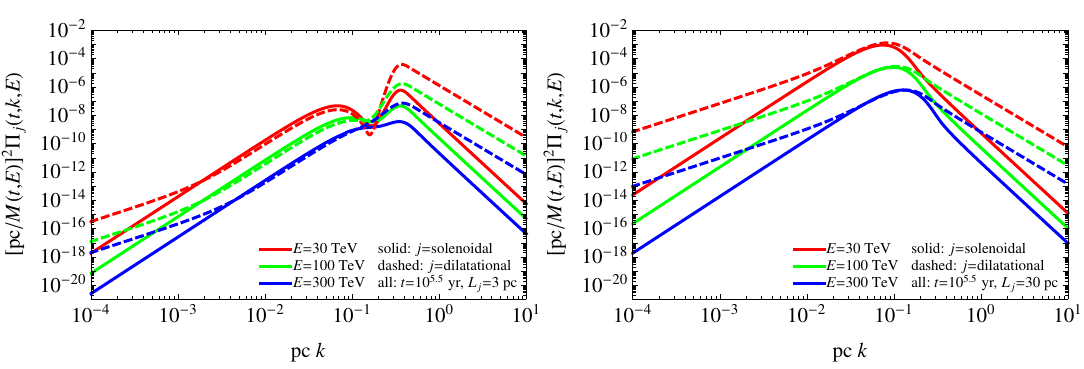}
\caption{Fluctuation power spectra of the gamma-ray halo around a Geminga-like source.}\label{rad}
\end{figure*}

In the intermediate region between the large- and small-scale limits, the spectral shape of the halo largely depends on the relative magnitude of $ \Lambda $ and $ L_j $, which is evident in Figure \ref{a&Lambda}. Owing to the cutoff of $ w_j\left( k\right) $ for $ k<1/L_j $, there may be a dip in $ P_j\left( t,k,p\right) $ in $ 1/\Lambda\left( t,p \right) <k<1/L_j $ if $ \Lambda\left( t,p \right) >L_j $, as shown by some of the lines in the left-hand panels of Figures \ref{modSNR} and \ref{modPWN}. While other lines indicate that $ P_j\left( t,k,p\right) $ can have a smooth break around $ k\sim 1/\Lambda\left( t,p \right) $ if $ \Lambda\left( t,p \right) <L_j $.
\section{Radiation Power Spectrum}
Although we have approximated the total loss rate with Equation \eqref{b}, the detailed KN effect may not be omitted when calculating the emissivity at a given scattered photon energy $ E $, especially for TeV gamma rays produced by interstellar electrons and positrons. This production is governed by the direction-integrated KN cross section in the ultra-relativistic limit,
\begin{align}
\frac{d\sigma _{\text{IC}}\left( p,E,\varepsilon \right)}{dE}&=\frac{3\sigma _{\text{T}}}{4\varepsilon}\left( \frac{mc}{p} \right) ^2F\left( p,E,\varepsilon \right) \notag\\
&\hspace{-5em}\times \Theta \left[ q\left( p,E,\varepsilon \right) -\frac{1}{4}\left( \frac{mc}{p} \right) ^2 \right] \Theta \left[ 1-q\left( p,E,\varepsilon \right) \right] 
\end{align}
\citep{1970RvMP...42..237B}, where $ \varepsilon $ is the initial photon energy in the scattering process, and
\begin{align}
F\left( p,E,\varepsilon \right) &=2q\left( p,E,\varepsilon \right) \ln q\left( p,E,\varepsilon \right) +\left[ 1-q\left( p,E,\varepsilon \right) \right] \notag\\
&\hspace{-4em}\times\left\{ 1+2q\left( p,E,\varepsilon \right) +\frac{1}{2}\frac{\left[ \Gamma _\varepsilon\left( p \right) q\left( p,E,\varepsilon \right) \right] ^2}{1+\Gamma _\varepsilon\left( p \right) q\left( p,E,\varepsilon \right)} \right\} ,\\
q\left( p,E,\varepsilon \right) &=\frac{1}{\Gamma _\varepsilon\left( p \right)}\frac{E}{cp-E},\quad\Gamma _\varepsilon\left( p \right) =\frac{4\varepsilon}{mc^2}\frac{p}{mc}.
\end{align}

The distribution of the IC scattered photons is
\begin{align}
\phi\left( t,\boldsymbol{r},E \right) &=4\pi c\int_0^{\infty}\int_0^{\infty}\frac{d\sigma _{\text{IC}}\left( p,E,\varepsilon \right)}{dE}\notag\\
&\quad\times n\left( \varepsilon \right) f\left( t,\boldsymbol{r},p \right) p^2dpd\varepsilon ,\label{phi}
\end{align}
where $ n $ is the seed photon distribution, typically characterized as a black-body spectrum. Owing to convective oscillations, $ \phi $ fluctuates with the fluctuations of the CR distribution $ f $. This renders $ \phi =\phi _0+\phi _1 $, where $ \phi _0 $ and $ \phi _1 $ are determined by $ f_0 $ and $ f_1 $, respectively. It follows from Equations \eqref{f0} and \eqref{f0k} that $ \phi _0 $ and its Fourier transform $ \tilde{\phi}_0 $ are Gaussian-like, with the amplitude in Fourier space evaluated as
\begin{align}
M\left( t,E \right) &=4\pi c\int_0^{\infty}\int_0^{\infty}\frac{d\sigma _{\text{IC}}\left( p,E,\varepsilon \right)}{dE}\notag\\
&\quad\times n\left( \varepsilon \right) a\left( t,p \right) p^2dpd\varepsilon .
\end{align}

We define the power spectrum of the random field $ \phi _1 $ as
\begin{align}
\Pi\left( t,\boldsymbol{k},E \right) =4\pi k^2\left< \tilde{\phi}_1\left( t,\boldsymbol{k},E \right) \tilde{\phi}_{1}^{*}\left( t,\boldsymbol{k} ,E \right) \right> ,
\end{align}
where $ \tilde{\phi}_1 $ is the Fourier transform of $ \phi _1 $. With Equations \eqref{f1kf1k'}, \eqref{Ptlk}, and \eqref{phi}, the component spectrum is given by
\begin{align}
\Pi _j\left( t,k,E \right) &=\left( 4\pi c \right) ^2\notag\\
&\hspace{-3em}\times\int_0^{\infty}\int_0^{\infty}\int_0^{\infty}\int_0^{\infty}\frac{d\sigma _{\text{IC}}\left( p,E,\varepsilon \right)}{dE}\frac{d\sigma _{\text{IC}}\left( p' ,E,\varepsilon ' \right)}{dE}\notag\\
&\hspace{-3em}\times n\left( \varepsilon \right) n\left( \varepsilon ' \right) \mathcal{P} _j\left( t;k;p,p' \right) \left( pp' \right) ^2dpdp'd\varepsilon d\varepsilon '.\label{Pi}
\end{align}
From the rule-of-thumb formulae for IC energy scaling, we expect that the $ k $-dependences of $ \Pi _j\left( t,k,E \right) $ in the Thomson and extreme KN limits mimic those of $ P_j\left( t,k,mc\sqrt{E/\varepsilon} \right) $ and $ P_j\left( t,k,E/c \right) $, respectively.

Assuming the seed photons are from the CMB, in Figure \ref{rad} we display some numerical results of Equation \eqref{Pi} for the model of Equations \eqref{aPWN} and \eqref{LambdaPWN} with $ t=10^{5.5} $ yr, which corresponds to a Geminga-like source \citep{2017Sci...358..911A}. Comparing Figures \ref{modPWN} and \ref{rad}, we find that the spectral indices of $ \Pi _j $ are asymptotic to those of $ P_j $ in the large- and small-scale limits; moreover, $ \Pi _j\left( t,k,30\text{ TeV} \right) $ well mimics $ P_j\left( t,k,100\text{ TeV}/c \right) $ for any $ k $. This implies that the IC radiation power spectra directly reveal the global shapes of the electron and positron power spectra near PWNe.

However, the 3D radiation distribution $ \phi $ remote from the Earth is not accessible in existing observations. The direct observable is the surface brightness
\begin{align}
S\left( t,\boldsymbol{R},E \right) =\frac{E}{4\pi}\int_0^{\infty}{\phi \left( t,\boldsymbol{Z}-\boldsymbol{Z}_{\text{PWN}},E \right) dZ},
\end{align}
where $ \boldsymbol{Z} $ and $ \boldsymbol{Z}_\text{PWN} $ are the vectors from the Earth to the field point and PWN, respectively, and $ \boldsymbol{R} $ is the position vector along the celestial sphere. Since most emitted photons are distributed within a small region around the PWN, the lower limit of the line-of-sight integral can be extended to $ -\infty $, and the projection-slice theorem becomes approximately valid. Thus, the Fourier transform of $ S\left( t,\boldsymbol{R},E \right) $ is
\begin{align}
\tilde{S}\left( t,\boldsymbol{\mathcal{K}},E \right) =\frac{E}{4\pi}\tilde{\phi}\left( t,\boldsymbol{\mathcal{K}},E \right) e^{-i\boldsymbol{\mathcal{K}}\cdot \boldsymbol{Z}_{\text{PWN}}},
\end{align}
where $ \boldsymbol{\mathcal{K}} $ is the wave vector along the celestial sphere. This means that the surface-brightness fluctuation power spectrum $ 2\pi \mathcal{K}\left< \tilde{S}_1\left( t,\boldsymbol{\mathcal{K}},E \right) \tilde{S}_{1}^{*}\left( t,\boldsymbol{\mathcal{K}},E \right) \right> $ scales as $ \Pi \left( t,\mathcal{K},E \right) /\mathcal{K} $.
\section{Summary}
In the scenario of convection--diffusion propagation, we demonstrated that CRs can have convective oscillations forced by turbulent convection. Our main findings are as follows:
\begin{enumerate}
\item Under the assumption that the convection field is statistically homogeneous and isotropic, and that the ensemble-averaged distribution of CRs is spherically symmetric, the spatial fluctuations of the distribution form an inhomogeneous isotropic random field, which can be decomposed into incompressible and acoustic components.
\item The Fourier power of these components roughly follows peaked broken power-law spectra, whose asymptotic spectral indices in the small-scale limit (Equation \ref{SCL}) depend on the slopes of the velocity power spectra, and those in the large-scale limit (Equation \ref{LSL}) arise purely from the structure of Parker's equation.
\item Convective oscillations of interstellar electrons and positrons can generate a random halo of TeV photons in the IC scattering process, and the power spectrum of such a gamma-ray halo around a PWN-like source should simply mimic that of the seed leptonic halo.
\end{enumerate}

These points provide a feasible framework to describe the irregular structures of the TeV gamma-ray halos imaged around PWNe. We expect that the power spectra of the gamma-ray images can be used to probe the properties of local interstellar turbulence on scales below the extended source size. Fourier or structure-function analyses for future high-resolution gamma-ray images are necessary to extract the power spectra.

We emphasize that the present model is based on the assumption that the ensemble-averaged CR distribution is Gaussian. For PWN halos, this assumption may be sufficient to analyze the large- and small-scale limits of the power spectra, but too coarse to derive all intermediate states between the two limits. Another limitation arises from the absence of anisotropic diffusion and statistically anisotropic convection. These anisotropies may result from a coherent magnetic field \citep{2022MNRAS.511.6218Z,2025ApJ...988...20Z}, and are important when the observed PWN halo exhibits strong systematic asymmetry \citep{2024ApJ...976....4D,2025arXiv251006786C}. Further work is warranted.
\section*{Acknowledgments}
This work is supported by grants from the National Natural Science Foundation of China (Nos.~12303055, 12375103), and the Fundamental Research Funds for the Central Universities (2682025CX030).
\bibliography{ref}
\bibliographystyle{aasjournal}
\end{document}